\documentclass[letterpaper]{article} 
\usepackage{aaai25}  
\usepackage{times}  
\usepackage{helvet}  
\usepackage{courier}  
\usepackage[hyphens]{url}  
\usepackage{graphicx} 
\usepackage{natbib}  
\usepackage{caption} 
\usepackage{algorithm}
\usepackage{algorithmic}

\usepackage{newfloat}
\usepackage{listings}
\DeclareCaptionStyle{ruled}{labelfont=normalfont,labelsep=colon,strut=off} 
\floatstyle{ruled}
\newfloat{listing}{tb}{lst}{}
\floatname{listing}{Listing}
\usepackage{cuted} 
\usepackage{multirow}
\usepackage{amsmath}
\usepackage{amsfonts}
\usepackage{amsthm}
\usepackage{amssymb}
\usepackage{makecell}
\usepackage{booktabs}
\usepackage{subcaption,xcolor,colortbl}

\title{VRVVC: Variable-Rate NeRF-Based Volumetric Video Compression}
\author {
    Qiang Hu\textsuperscript{\rm 1}\equalcontrib,
    Houqiang Zhong\textsuperscript{\rm 2}\equalcontrib,
    Zihan Zheng\textsuperscript{\rm 1},
    Xiaoyun Zhang\textsuperscript{\rm 1}\thanks{Corresponding author.},
    Zhengxue Cheng\textsuperscript{\rm 2},
    Li Song\textsuperscript{\rm 2},
    Guangtao Zhai\textsuperscript{\rm 2},
    Yanfeng Wang\textsuperscript{\rm 3}
}
\affiliations {
    \textsuperscript{\rm 1}Cooperative Medianet Innovation Center, Shanghai Jiao Tong University \\
    \textsuperscript{\rm 2}School of Electronic Information and Electrical Engineering, Shanghai Jiao Tong University \\
    \textsuperscript{\rm 3}School of Artificial Intelligence, Shanghai Jiao Tong University \\
    \{qiang.hu,zhonghouqiang,1364406834,xiaoyun.zhang,zxcheng,song\_li,zhaiguangtao,wangyanfeng\}@sjtu.edu.cn
}

\begin{document}

\maketitle

\begin{strip}\centering
\vspace{-24mm}
	\includegraphics[width=\linewidth]{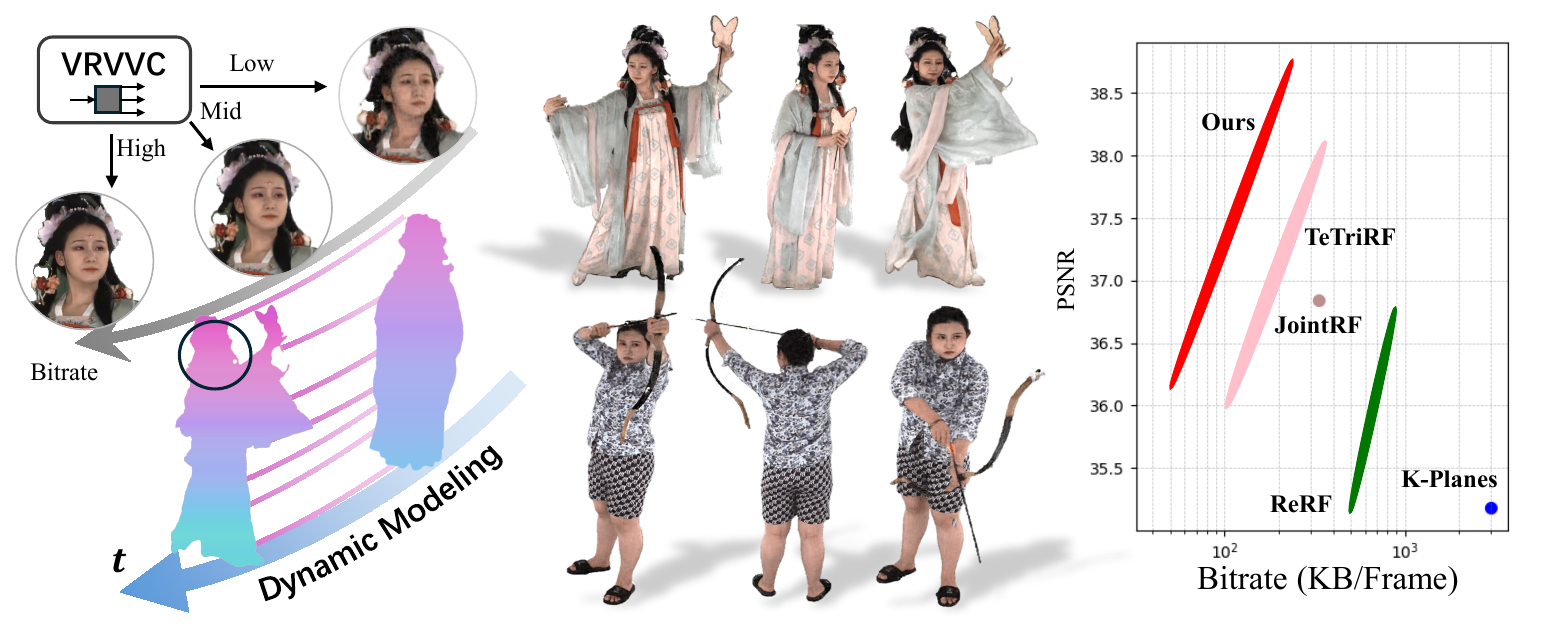}
	\vspace{-0.8cm}
	\captionof{figure}{\textbf{Left}: Our proposed VRVVC efficiently compresses volumetric video at variable bitrates using a single model. \textbf{Middle}: We demonstrate two examples of reconstruction quality at a bitrate of 60 KB per frame.  \textbf{Right}: The RD performance of our approach surpasses prior work (e.g. ReRF \cite{rerf}, TeTriRF \cite{tetrirf})} 
	\label{fig:teaser}
	\vspace{-5mm}
\end{strip}

\begin{abstract}
Neural Radiance Field (NeRF)-based volumetric video has revolutionized visual media by delivering photorealistic Free-Viewpoint Video (FVV) experiences that provide audiences with unprecedented immersion and interactivity. However, the substantial data volumes pose significant challenges for storage and transmission. Existing solutions typically optimize NeRF representation and compression independently or focus on a single fixed rate-distortion (RD) tradeoff. In this paper, we propose VRVVC, a novel end-to-end joint optimization variable-rate framework for volumetric video compression that achieves variable bitrates using a single model while maintaining superior RD performance. Specifically, VRVVC introduces a compact tri-plane implicit residual representation for inter-frame modeling of long-duration dynamic scenes, effectively reducing temporal redundancy. We further propose a variable-rate residual representation compression scheme that leverages a learnable quantization and a tiny MLP-based entropy model. This approach enables variable bitrates through the utilization of predefined Lagrange multipliers to manage the quantization error of all latent representations. Finally, we present an end-to-end progressive training strategy combined with a multi-rate-distortion loss function to optimize the entire framework. Extensive experiments demonstrate that VRVVC achieves a wide range of variable bitrates within a single model and surpasses the RD performance of existing methods across various datasets.
\end{abstract}

\section{Introduction}
Photorealistic volumetric video provides an immersive experience in virtual reality and telepresence, demonstrating significant potential to become the next-generation video format. Traditional approaches to volumetric video reconstruction have primarily relied on point cloud-based methods \cite{graziosi2020overview} and depth-based techniques \cite{boyce2021mpeg}, which often struggle with rendering quality and realism. Recently, both Neural Radiance Fields (NeRF) \cite{mildenhall2021nerf} and 3D Gaussian Splatting (3DGS) \cite{kerbl20233d} have shown considerable promise in representing photorealistic volumetric video. However, challenges remain in the storage and transmission of volumetric video with NeRF or 3DGS. Compared to the implicit modeling of NeRF, 3DGS utilizes an explicit point cloud representation, which is less conducive to efficient compression. In summary, NeRF's compact representation and implicit modeling capabilities make it inherently suitable for volumetric video compression.

NeRF and its variants \cite{instant-ngp,merf} have achieved remarkable success in synthesizing novel views, inspiring a multitude of derivative research studies focused on dynamic scenes. Some techniques \cite{Nerfies,DNeRF} employ deformation fields to capture voxel movements relative to a canonical space, while others \cite{tineuvox,HumanRF,kplanes} introduce temporal voxel features or apply joint training across multiple frames to achieve superior temporal reconstructions. However, most existing studies primarily focus on improving the reconstruction quality of NeRF representations, frequently neglecting the critical need to minimize storage size and transmission bandwidth. This oversight poses substantial challenges for practical applications, especially in streaming volumetric video.


To address these problems, several approaches have been proposed to compress explicit features of dynamic NeRF. For instance, ReRF \cite{rerf} uses a grid-based explicit representation to model the spatial-temporal feature space of dynamic scenes and adopts traditional image encoding techniques to compress the representation after training. TeTriRF \cite{tetrirf} utilizes a hybrid representation with tri-plane to model dynamic scenes and employs a traditional video codec to further reduce redundancy. However, these methods optimize representation and compression independently, neglecting the rate-distortion (RD) tradeoff during the training phase, which ultimately limits their compression performance. To close this gap, JointRF \cite{zheng2024jointrf} introduces an end-to-end combined training approach for dynamic NeRF representation and compression, but it is fixed-rate only and suffers from a slow rendering speed.


In this paper, we propose VRVVC, a novel variable-rate compression framework tailored for NeRF-based volumetric video. Our key idea involves estimating the bitrate of NeRF representations during end-to-end training and controlling it using the RD tradeoff parameter $\lambda$. By incorporating both bitrate and distortion terms into the loss function, we achieve optimal RD performance across a wide range of variable bitrates using a single model, as illustrated in Fig. \ref{fig:teaser}. We realize this through three main innovations. First, we introduce a compact tri-plane implicit residual representation for inter-frame modeling within long sequences.  For each frame, VRVVC decomposes the radiance field into a tri-plane and models the residual information between adjacent timestamps within this feature space. This representation effectively captures high-dimensional appearance features within compact planes. 

Second, we propose a variable-rate residual representation compression scheme that leverages a learnable quantization step and a tiny MLP-based entropy model, combined with a predefined set of Lagrange multipliers, to facilitate variable bitrates. Third, we present an end-to-end progressive learning scheme to jointly optimize both the representation and compression. This approach yields temporally consistent and low-entropy 4D sequential representations that can be effectively compressed, significantly enhancing RD performance. Experimental results show that our VRVVC achieves variable bitrates by a single model while maintaining state-of-the-art RD performance across various datasets. Compared to the previous leading method, TeTriRF \cite{tetrirf}, our approach achieves approximately \textbf{-81\%} BD-rate savings on the DNA-Rendering \cite{dna} dataset and an \textbf{-46\%} BD-rate reduction on the ReRF dataset.


In summary, our contributions are as follows:
\begin{itemize}    
    \item We propose VRVVC, a novel approach for variable-rate compression of NeRF-based volumetric video. Our VRVVC achieves variable bitrates within a single model while delivering improved RD performance.    
    \item We introduce a compact and compression-friendly representation that models volumetric video as a tri-plane residual radiance field, effectively minimizing temporal redundancy for inter-frame modeling of extended dynamic scenes.
    \item We present an end-to-end progressive training scheme that jointly optimizes representation and compression through a multi-rate-distortion loss function, significantly improving compression performance compared to post-training methods.
\end{itemize}

\section{Related Work}
\textbf{Dynamic Radiance Field Representation.} 
NeRF \cite{mildenhall2021nerf} employs implicit representations to synthesize highly realistic novel views. Its advancements \cite{instant-ngp,rabich2024fpo++,martinbrualla2020nerfw,barron2021mipnerf,barron2022mipnerf360} in static scenes have catalyzed research into dynamic scenes, particularly in volumetric video. Deformation field techniques \cite{NeuralRadianceFlow,li2022neural,DNeRF,nerfplayer} recover temporal features by warping real-time frames to a canonical space. However, these methods struggle with large-scale motions and changes, resulting in slower training and rendering. Conversely, other  approaches \cite{tineuvox,HumanRF,kplanes,hexplane_2023_CVPR,streaming,tensor4d} extend the radiance field into a 4D spatio-temporal domain, facilitating faster training and rendering 
at the cost of increased storage demands. 
Several studies \cite{rerf,videorf,tetrirf,zheng2024jointrf,10.1145/3664647.3681107} use residual radiance fields to represent long-sequence dynamic scenes, leveraging compact motion grids and residual feature grids to exploit inter-frame feature similarity. 
Our compact tri-plane residual-based dynamic modeling method is designed for inter-frame modeling in extended sequences, which effectively captures high-dimensional appearance features within compact planes.

\textbf{NeRF Compression.} 
Recently, deep learning-based image and video compression methods have demonstrated strong RD performance for 2D video \cite{10632166,balle2016end,balle2018variational,2017endtoend,9150865,Choi_2019_ICCV,9578818,06b2c0bf3a91491aa41a528bba116a14,9879179}. Efforts are now being made to extend these compression techniques to the NeRF domain \cite{vqrf,ecrf,peng2023representing,miniwave}. VQRF \cite{vqrf} and ECRF \cite{ecrf} have made strides by employing entropy encoding and frequency domain mapping, respectively, for compressing static radiance fields. However, these methods are limited to static scenes and do not address dynamic scenarios.
Recent studies like ReRF, VideoRF \cite{videorf}, and TeTriRF \cite{tetrirf} focus on dynamic scenes. They integrate traditional image and video encoding techniques for feature compression but fail to jointly optimize the representation and compression of the radiance field, resulting in a loss of dynamic details and compression efficiency.
Our approach estimates the bitrate of representations during training and controls it using the RD tradeoff parameter $\lambda$, enabling end-to-end training. This allows our model to achieve a wide range of variable bitrates, unlike JointRF, which is restricted to a fixed bitrate.

\section{Method}
\begin{figure}[ht]
\vspace{-3mm}
    \centering
    \includegraphics[width=\linewidth]{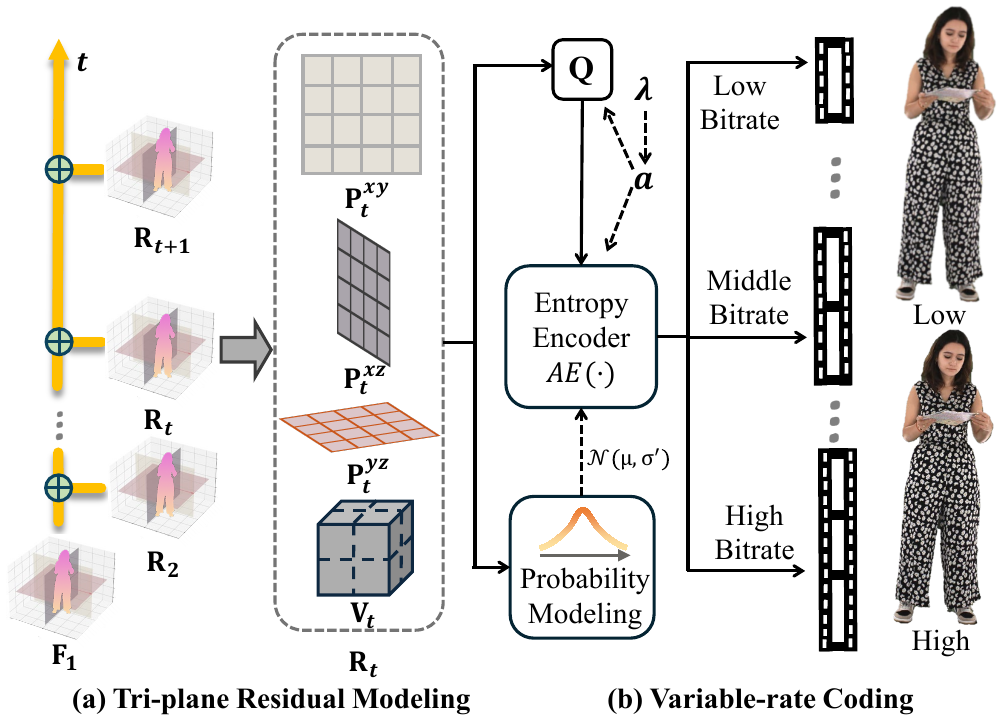}
    \vspace{-6mm}
    \caption{Illustration of our VRVVC framework. We employ a compact tri-plane residual representation for inter-frame modeling of long-duration dynamic scenes. The residuals are encoded into several bitstreams in an MLP-based entropy model that utilizes the RD tradeoff parameter $\lambda$ to achieve variable bitrates within a single model.  }
    \label{fig:overview}
\end{figure}

In this section, we introduce the details of the proposed VRVVC. Fig. \ref{fig:overview} illustrates the overall framework of our method. We model the inter-frame relationships of long dynamic scenes using a compact tri-plane residual representation. Additionally, we propose a variable-rate entropy coding scheme to achieve a wide range of variable bitrates within a single model. We also introduce a fast progressive training strategy that jointly optimizes representation and compression, greatly improving compression efficiency while preserving high rendering quality.

\vspace{-4mm}
\subsection{Tri-plane Residual Dynamic Modeling} \label{3.1}
Recall that a NeRF models a 3D volumetric scene using a 5D function $\Psi$, which maps the spatial coordinate $\mathbf{x} = (x, y, z)$ and view direction $\mathbf{d} = (\theta, \phi)$ to color $\mathbf{c}$ and density $\sigma$, formulated as $(\mathbf{c}, \sigma) = \Psi(\mathbf{x}, \mathbf{d})$. Then, volume rendering is employed for photo-realistic novel view synthesis. To enhance training and rendering efficiency, we employ a feature tri-plane $\mathbf{P} = \{ \mathbf{P}^{l} \mid l \in L \}$, $ L = \{ xy, yz, xz\}$ along with a 3D density grid $\mathbf{V}$ as our static representation $\mathbf{F} = ( \mathbf{P}, \mathbf{V})$. Specifically, the radiance field of a static scene is: 
\vspace{-1mm}
\begin{equation}
    \begin{aligned}
        \mathbf{\mathbf{f}} &= \bigcap_{l \in L } \varphi \left( \pi_l (\mathbf{x}, \mathbf{P}^l) \right)  \\
        \mathbf{c} &= \Phi (\mathbf{\mathbf{f}}, \omega (\mathbf{d})) \\
        \sigma &= \varphi \left(\mathbf{x}, \mathbf{V} \right)
    \end{aligned}
    \vspace{-1mm}
\end{equation}
where $\varphi$ denotes the interpolation function, $\pi_l$ projects the 3D point $\mathbf{x}$ onto feature plane $l$, and $\bigcap$ represents concatenating the features from three planes. The MLP $\Phi$ decodes the color at point $\mathbf{x}$ based on the concatenated feature $\mathbf{f}$ and the encoded view direction $\omega(\mathbf{d})$. The density of point $\mathbf{x}$ is derived through interpolation on the density grid.

When expanding from static to dynamic scenes, a straightforward approach is to utilize individual per-frame features to represent a dynamic scene composed of $M$ frames, denoted as $\{\mathbf{F}_t\}_{t=1}^M$. However, this approach neglects temporal coherence, resulting in substantial temporal redundancy. Conversely, other methods \cite{Cao_2023_CVPR,kplanes} that directly model entire dynamic scenes using NeRF representation may lead to suboptimal performance for long sequences and are unsuitable for streaming applications. To address these challenges, we extend the current static NeRF representation to dynamic scenes by employing a frame-by-frame tri-plane residual inter-frame modeling strategy. 

Our tri-plane residual modeling method divides the entire sequence into equal-length groups of features (GoFs), each containing $N$ frames. In each GoF, the first frame serves as an I-feature $\mathbf{F}_1$ (keyframe) that is modeled independently, while the subsequent frames are designated as P-features $\{\mathbf{R}_t\}_{t=2}^N$, which represent the compensated residual relative to the preceding feature. Besides, frames within the same group share a compact global MLP $\Phi$ as the feature decoder for the spatial-temporal feature space, effectively reducing bitrate consumption while maintaining performance quality. Finally, our VRVVC sequentially represents a GoF with N frames as $\Phi$ and $\mathbf{G} = \{ \mathbf{F}_{1}, \mathbf{R}_{2}, \mathbf{R}_{3} \cdots \mathbf{R}_{t}  \cdots \mathbf{R}_{N} \}$, as illustrated in Fig.  \ref{fig:overview}.

\begin{figure*}[ht]
\vspace{-3mm}
    \centering
    \includegraphics[width=\linewidth]{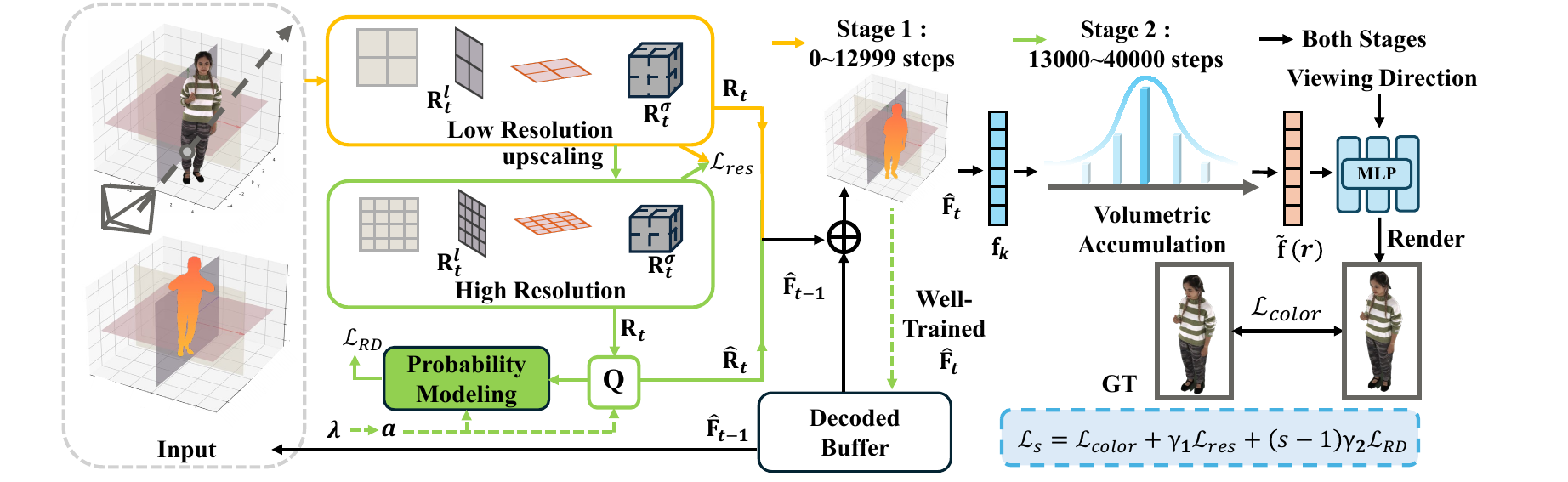}
    \vspace{-6mm}
    \caption{Overview of our progressive training. In the first stage, we adopt the reconstructed features $\hat{\mathbf{F}}_{t-1}$ from the previous frame, retrieved from the decoded buffer, to train the current frame's low-resolution residual features. In the second stage, these features are reused as an effective initialization for further training, where they are integrated with a variable-rate entropy coding model for joint optimization. The entire training process is supervised by the multi-rate-distortion loss $\mathcal{L}_s$.}
    \label{fig:train}
    \vspace{-3mm}
\end{figure*}

Our VRVVC enables highly efficient sequential modeling of P-features by leveraging inter-frame feature similarities. Specifically, we retrieve the reconstructed feature of the previous frame $\hat{\mathbf{F}}_{t-1}$ from the decoded buffer and combine it with the input images of the current frame to learn the residual for the current frame $\mathbf{R}_t$, as shown in Fig. \ref{fig:train}. Then, we can reconstruct the entire feature of the current frame $\hat{\mathbf{F}}_{t}$ by applying the residual compensation:
\begin{equation}
\begin{aligned}
    \hat{\mathbf{F}}_t &= \hat{\mathbf{F}}_{t-1} + \hat{\mathbf{R}}_t \\
    &= (\bigcup_{l \in L} ( \hat{\mathbf{P}}_{t-1}^l + \hat{\mathbf{R}}^{l}_{t} ),  \hat{\mathbf{V}}_{t-1} + \hat{\mathbf{R}}^{\sigma}_{t})
\end{aligned}
\vspace{-3mm}
\end{equation}
where $\bigcup$ represents the union of tri-plane features, and $\hat{\mathbf{R}}_t=\{ \hat{\mathbf{R}}^{xy}_{t}, \hat{\mathbf{R}}^{yz}_{t},\hat{\mathbf{R}}^{xz}_{t},\hat{\mathbf{R}}^{\sigma}_{t} \}$ denotes the reconstruction residual for tri-plane and density grid. Finally, $\hat{\mathbf{F}}_{t}$ is stored in the decode buffer for the reconstruction of the next frame. 

It is worth noting that our tri-plane residual representation has several significant advantages. Firstly, it is both effective and compact, capable of capturing high-dimensional appearance features and decomposing them into three orthogonal feature planes. Secondly, it is highly compression-friendly, as it leverages the simplicity of the residual data distribution to efficiently reduce spatio-temporal redundancy between frames. Thirdly, it facilitates efficient training and rendering by incorporating an explicit density grid, which enables rapid retrieval of density values. This allows for the swift removal of sample points in empty space without the need for network inference, thus accelerating both training and inference processes.

\subsection{Variable-rate Entropy Coding}
We also propose a variable-rate entropy coding scheme for residual representation, enabling flexible adjustments between different bitrates and reconstruction quality within a single model. Unlike traditional methods \cite{yang2020variable,lin2021deeply} that adjust the interval of the fixed Lagrange multiplier in universal quantization, our method integrates $\lambda$ with a univariate quantization regulator $a$ to control the quantization error of the overall latent representation, achieving a wide range of variable bitrates.

A shared two-layer MLP is first used to extract high-dimensional latent representation $\mathbf{y}_t$ from the residual $\mathbf{R}_t$, aggregating feature information while mitigating compression-induced information loss. This is followed by a CNN with five 3x3 layers, which refines the features and generates the final context feature $\mathbf{z}_t$.
We then estimate the Gaussian entropy $p(\mathbf{\hat{y}}_t|\mathbf{\hat{z}}_t)$ of the quantized latent representation $\hat{\mathbf{y}}_t$ on condition of quantized context feature $\hat{\mathbf{z}}_t$ . 
This estimation guides the arithmetic entropy coding of $\hat{\mathbf{y}}_t$ into a bitstream. 
In this paper, we use a tiny MLP to predict $p(\mathbf{\hat{y}}_t|\mathbf{\hat{z}}_t)$ as follows:
\begin{align}
    p(\mathbf{\hat{y}}_t|\mathbf{\hat{z}}_t) = \mathcal{N}(\mu, \sigma^{\prime}) * \mathcal{U}(-\frac{1}{2}, \frac{1}{2}) (\mathbf{\hat{y}}_t) \label{equ:entropy}
\end{align}
where $\mathcal{N}(\mu, \sigma^{\prime})$ denotes the Gaussian distribution. 

When $\mathbf{y}_t$ undergoes different quantization operations, its probability distribution can vary significantly, leading to substantial quantization errors. To mitigate this, we introduce a set of learnable quantization parameters $\mathbf{A} = \{a_1, a_2, \dots, a_n\}$, coupled with predefined Lagrange multipliers $\Lambda = \{ \lambda_1, \lambda_2 \cdots \lambda_n \}$ to control these errors and enable variable bitrates. The learnable quantization parameter $a_i$ adjusts the quantization bin size and impacts bitrate, while the Lagrange multiplier $\lambda_i$ controls the trade-off between bitrate and distortion, creating a coupling relationship between $a_i$ and $\lambda_i$. In learning-based image codecs, $\sqrt{\lambda_i}$ is nearly proportional to $a_i$, whereas in video codecs, QP is proportional to $\text{ln}(\lambda)$. Thus, pairing $a_i$ with $\lambda_i$ better balances the RD trade-off, and the values are averaged across different scenes to ensure broad applicability.
The latent representation $\mathbf{y}_{t}$ is initially scaled by its corresponding parameter $a_i$ before being quantized into $\hat{\mathbf{y}}_{t}$ as follows:
\begin{align}
\hat{\mathbf{y}}_t = \text{round}\left(\frac{\mathbf{y}_t}{a_i}\right) \cdot a_i , \quad a_i \in \mathbf{A}. \label{equ:quant}
\end{align}
The entropy model of $\hat{\mathbf{y}}_t$ in Eq. \ref{equ:entropy} is then rewritten as:
\begin{align}
    p(\hat{\mathbf{y}}_t|\hat{\mathbf{z}}_t,a_i) = \mathcal{N}(\mu_i, \sigma_i^{\prime}) * \mathcal{U}\left(-\frac{1}{2a_i}, \frac{1}{2a_i}\right)(\hat{\mathbf{y}}_t)
\end{align}

Since the quantization operation is inherently non-differentiable, we also apply a straight-through estimator (STE) to approximate the gradient during backpropagation. The STE facilitates gradient flow through the quantization step by approximating the gradient as $\frac{\partial \hat{\mathbf{y}}_t}{\partial \mathbf{y}_t} \approx 1$. This approximation enables effective optimization of the learnable quantization parameters $a_i$ during training, allowing the model to dynamically adjust the quantization step size and optimize the bitrate. 
The RD loss function for the variable-rate model is formulated as:
\begin{equation}
\begin{aligned}
\mathcal{L}_{RD} = \sum_{i=1}^n ( \mathbb{E} [-\log p(\hat{\mathbf{y}_t}|\hat{\mathbf{z}}_t,a_i)] + \lambda_i \cdot D(\mathbf{R}_t, \hat{\mathbf{R}}_t) )
\end{aligned}    
\end{equation}
where $\mathbb{E} [-\log p(\hat{\mathbf{y}_t}|\hat{\mathbf{z}}_t,a_i)]$ represents the estimated bitrate required to encode $\hat{\mathbf{y}}_t$, while $D(\mathbf{R}_t, \hat{\mathbf{R}}_t)$ measures the distortion between the original residual $\mathbf{R}_t$ and its reconstruction $\hat{\mathbf{R}}_t$. The Lagrange multiplier $\lambda_i$ paired with $a_i$ balances the trade-off between bitrate and distortion. Our approach integrates a univariate quantization regulator into the quantization and entropy coding process to control quantization error. By applying rate-distortion supervision across various quantization parameters, we achieve variable bitrates within a single model.

\subsection{Progressive Training Strategy}
Here, we introduce an end-to-end progressive training scheme that jointly optimizes both the representation and compression to further improve RD performance. An overview of our progressive training, which incorporates a two-stage coarse-to-fine strategy, is illustrated in Fig. \ref{fig:train}. In the first stage, we train the density grid and feature planes at a low resolution, enabling rapid exploration of the scene's core structure. In the second stage, we leverage the low-resolution feature planes from the first stage as an effective initialization for subsequent training, combining them with a variable-rate entropy coding model for joint optimization. Our approach dramatically accelerates training while improving both rendering quality and compression efficiency.

\textbf{Stage 1.}
The inputs for this stage include the multi-view images of the current frame and the reconstructed features $\hat{\mathbf{F}}_{t-1}$ of the previous frame obtained from the decoded buffer. These reconstructed features are downsampled to a low resolution, serving as the initialization for the coarse training stage.
The outputs of this stage are the residual tri-plane features and the density grid, both at a low resolution. The density grid provides a rough approximation of the scene's geometry, which is essential for identifying and eliminating idle spaces during the preliminary reconstruction of the density field, thereby reducing unnecessary computational overhead. This training stage not only accelerates convergence but also establishes a solid foundation for more detailed optimization in stage 2.

\textbf{Stage 2.}
The residual features generated in stage 1 are upsampled to a higher resolution and used as initialization for the second training stage. By reusing these features instead of starting from scratch, we greatly reduce the training time and enhance convergence speed. Additionally, we employ a learnable variable-rate entropy coding model that is jointly trained with the residual dynamic modeling. During training, multiple $\lambda$ are used within the entropy model to optimize the quantization parameters $\mathbf{A}$, enabling variable-rate bitstreams. This joint training approach  effectively captures high-dimensional appearance features with low entropy, significantly enhancing compression efficiency while maintaining high rendering quality.

\textbf{Training Object.}
The multi-rate-distortion loss function of the entire framework is formulated as follows:
\begin{equation}
    \begin{aligned}
    \mathcal{L}_{s} = \mathcal{L}_{color} + \gamma_1 \mathcal{L}_{res} + (s-1)  \gamma_2 \mathcal{L}_{RD}, s \in \{1,2\}
\end{aligned}
\end{equation}
where $\mathcal{L}_{s}$ is the loss for stage $s$, $\gamma_1$ and $\gamma_2$ are the weights for our regular terms. $\mathcal{L}_{res} = \| \mathbf{R}_t \|_1$ serves as a residual regularization term, designed to ensure temporal continuity and minimize the magnitude of $\mathbf{R}_t$. $\mathcal{L}_{RD}$ represents variable-rate compression loss. $\mathcal{L}_{color}$ is the photometric loss,
\begin{equation}
    \begin{aligned}
\mathcal{L}_{color} =\sum_{r \in \Re} \| \mathbf{c}_g(\mathbf{r}) - \hat{\mathbf{c}}(\mathbf{r}) \|^2
\end{aligned}
\vspace{-2mm}
\end{equation}
where $\Re$ is the set of training pixel rays, $\mathbf{c}_g(\mathbf{r})$ and $\hat{\mathbf{c}}(\mathbf{r})$ are the ground truth and reconstructed colors of a ray $\mathbf{r}$, respectively.

\textbf{Rendering Acceleration.} 
In addition to utilizing a progressive training strategy, we also employ a deferred rendering model to further accelerate both the training and rendering processes. Specifically, We begin by accumulating the features along the ray:
\vspace{-2mm}
\begin{equation}
    \begin{aligned}
    \tilde{\mathbf{f}}(\mathbf{r}) &= \sum_{k=1}^{n_s} T_k \left(1 - \exp(-\sigma_k \delta_k)\right) \mathbf{f}_k, \\
    T_k &= \exp\left(-\sum_{j=1}^{k-1} \sigma_j \delta_j\right),
    \end{aligned}
    \vspace{-2mm}
\end{equation}
where $n_s$ represents the number of sample points along the ray $\mathbf{r}$, $\delta_k$ denotes the interval between adjacent samples. The density $\sigma_k = \varphi(\mathbf{k}, \mathbf{V})$ is interpolated from the density grid $\mathbf{V}$. We also leverage the density grid to eliminate points in empty space, thus reducing unnecessary computations. The composed feature $\mathbf{f}_k$ is formed by concatenating the appearance features $\mathbf{f}_k^l, l \in L$ from the tri-planes. The reconstructed color of the ray $\mathbf{r}$ is then computed using a tiny global MLP $\Phi$ that is shared across frames in the same GoF: $ \hat{\mathbf{c}}(\mathbf{r}) = \Phi(\tilde{\mathbf{\mathbf{f}}}(\mathbf{r}), \omega(\mathbf{d})) $
where $\omega(\mathbf{d})$ denotes the positional encoding of the viewing direction. This approach  significantly reduces computational complexity, as each ray requires only a single MLP decoding.

\begin{figure*}[ht]
\vspace{-4mm}
    \centering
    \includegraphics[width=\linewidth]{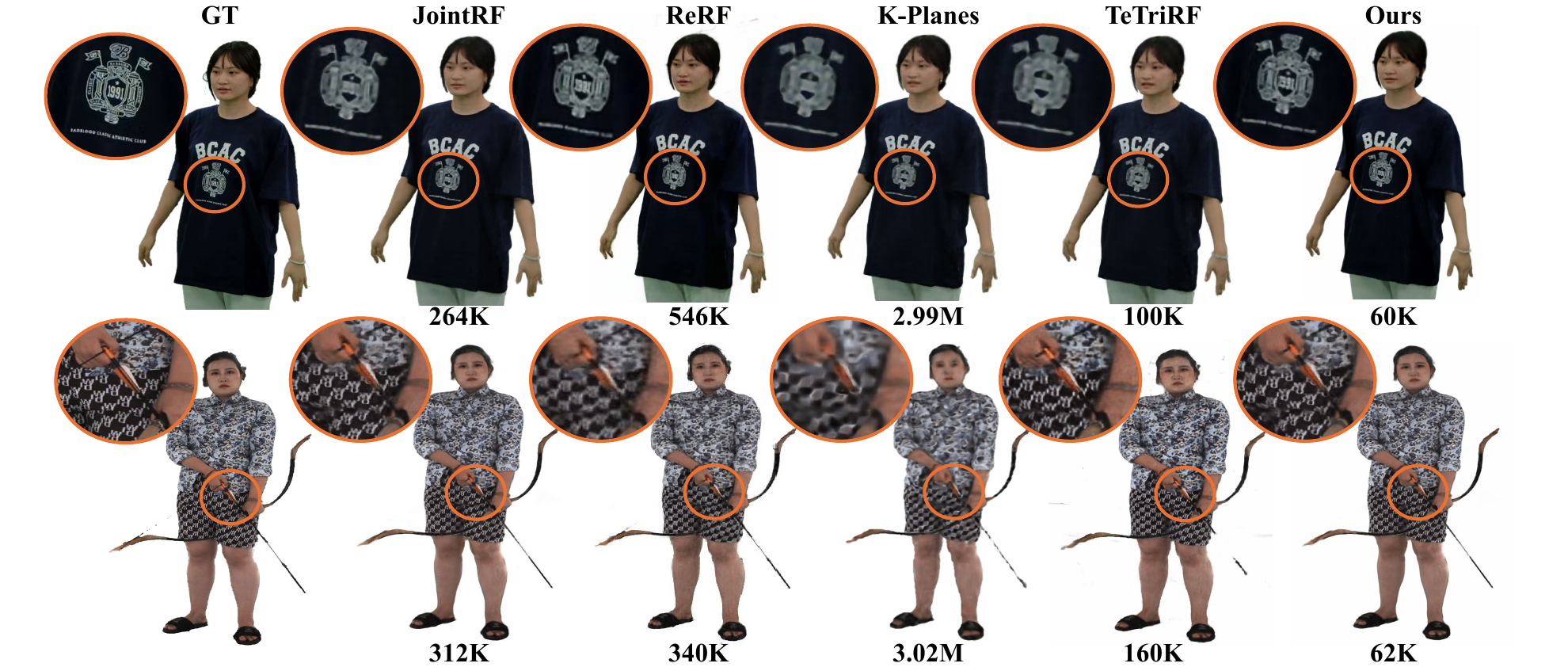}
    \vspace{-6mm}
    \caption{Qualitative comparison against volumetric video coding methods K-planes \cite{kplanes}, ReRF \cite{rerf}, TeTrirf \cite{tetrirf} and JointRF \cite{zheng2024jointrf}.}
    \label{fig:comparison}
    \vspace{-3mm}
\end{figure*}

\begin{table*}[htb]
\centering
\renewcommand\arraystretch{1.1}
\setlength{\tabcolsep}{4pt} 
\begin{tabular}{c|cccccc|cccccc}
\hline
\multicolumn{1}{l|}{} & \multicolumn{6}{c|}{ReRF Dataset}                                                                                                                     & \multicolumn{6}{c}{DNA-Rendering Dataset}                                                                                                                         \\ \hline
\multicolumn{1}{l|}{} & \multicolumn{2}{c}{Training View} & \multicolumn{2}{c|}{Testing View}                    & \multicolumn{1}{c|}{SIZE $\downarrow$} & R.T. $\downarrow$ & \multicolumn{2}{c}{Training View} & \multicolumn{2}{c|}{Testing View}                      & \multicolumn{1}{c|}{SIZE $\downarrow$} & R.T. $\downarrow$ \\ \cline{1-5} \cline{8-11}
Methods               & PSNR$\uparrow$  & SSIM$\uparrow$  & PSNR$\uparrow$ & \multicolumn{1}{c|}{SSIM$\uparrow$} & \multicolumn{1}{c|}{(KB)}                & (s)           & PSNR $\uparrow$ & SSIM $\uparrow$ & PSNR $\uparrow$ & \multicolumn{1}{c|}{SSIM $\uparrow$} & \multicolumn{1}{c|}{(KB)}                & (s)            \\ \hline
K-Planes              & 35.18           & 0.982           & 29.96          & 0.951                               & 2992                                   & 2.20               & 31.98           & 0.971           & 27.81           & 0.946                                & 3085                                   & 2.20              \\
ReRF                  & 35.20           & 0.982           & 30.88          & 0.962                               & 496                                    & 0.44              & 30.20           & 0.968          & 29.59           & 0.950                                & 314                                    & 0.47              \\
TeTriRF               & 35.94           & 0.986           & 32.05          & 0.974                               & 101                                   & 0.13              & 32.33           & 0.976           & 29.48           & 0.950                                & 160                                    & \textbf{0.12}     \\
JointRF               & 35.62           & 0.983           & 31.94          & 0.970                               & 227                                    & 0.78              & 33.74           & 0.979           & 30.27           & 0.962                                & 269                                    & 0.82              \\ \hline
Ours (Low)            & 35.93           & 0.986           & 32.16          & 0.975                               & \textbf{40}                            & \textbf{0.12}     & 33.24           & 0.978           & 30.25           & 0.962                                & \textbf{30}                            & \underline{0.13}        \\
Ours (Mid)            & \underline{36.31}           & \underline{0.988}     & \underline{32.45}    & \underline{0.976}                         & \underline{62}                                     & \textbf{0.12}     & \underline{34.45}     & \underline{0.981}     & \underline{31.37}     & \underline{0.968}                          & \underline{56}                                     & 0.14              \\
Ours (High)           & \textbf{38.73}  & \textbf{0.990}  & \textbf{33.52} & \textbf{0.978}                      & 223                                    & \underline{0.13}        & \textbf{35.95}  & \textbf{0.984}  & \textbf{32.45}  & \textbf{0.977}                       & 240                                    & 0.14              \\ \hline
\end{tabular}
\vspace{-2mm}
\caption{Quantitative comparison against volumetric video encoding methods. Bold data indicate the best performance, while underlined data indicate the second best.}
\label{t1}
\vspace{-3mm}
\end{table*}

\section{Experiment}
\subsection{Configurations}
\textbf{Datasets.} We validate the effectiveness of our method using two datasets: ReRF \cite{rerf} and DNA-Rendering \cite{dna}. The ReRF dataset consists of 74 camera views at a resolution of $1920 \times 2080$. We use 72 views for training and the remaining 2 for testing. Similarly, the DNA-Rendering dataset includes 48 views at a resolution of $2048 \times 2448$, with 46 views designated for training and the remaining 2 for testing. To ensure fairness across all comparative experiments, we specify the same bounding box for identical sequences.

\textbf{Setups.}
Our experimental setup includes an Intel E5-2699 v4 and a V100 GPU. We train 40,000 iterations, with each GoF lasting 30 frames. The Lagrange multipliers $\Lambda$ are initialized as \{0.0018, 0.0035, 0.0067, 0.0130, 0.025, 0.0483, 0.0932, 0.18\}, and the quantization parameters $A$ are set to \{1.0000, 1.3944, 1.9293, 2.6874, 3.7268, 5.1801, 7.1957, 10.0\}. The weights $\gamma_1$ and $\gamma_2$ are 0.0001 and 0.001, respectively.

\textbf{Evaluation Metrics.} To evaluate the compression performance of our method, we use Peak Signal-to-Noise Ratio (PSNR) and Structural Similarity Index (SSIM) \cite{1284395} as quality metrics. Bitrate is measured in KB per frame. For overall compression efficiency, we calculate the Bjontegaard Delta Bit-Rate (BDBR). Additionally, we assess rendering time efficiency (R.T.) by calculating the average rendering time per frame in seconds.

\subsection{Comparison}
We provide the experimental results to demonstrate the effectiveness of VRVCC. We compare with other state-of-the-art methods including K-planes \cite{kplanes}, ReRF \cite{rerf}, TeTrirf \cite{tetrirf} and JointRF \cite{zheng2024jointrf} both qualitatively and quantitatively. Fig. \ref{fig:comparison} displays qualitative comparisons for the \textit{kpop} sequence from ReRF and the \textit{Archer} sequence from DNA-Rendering. The results indicate that our VRVVC reconstructs finer details at a lower bitrate, such as the clothing in \textit{kpop} and the hand in \textit{Archer}, demonstrating the superior subjective experience provided by our method.

Tab.  \ref{t1} shows the detailed quantitative results on the ReRF and  DNA-Rendering datasets. In this table, ``Ours (Low)", ``Ours (Middle)" and ``Ours (High)" denote the performance of our method, which provides variable-rate bitstreams using a single model.  ``Ours (Middle)" achieves higher reconstruction quality at a lower bitrate compared to other methods. ``Ours (High)" offers significantly better reconstruction quality while requiring much less bitrate than K-planes and ReRF. ``Ours (Low)" achieves a bitrate substantially lower than TetriRF and JointRF with comparable reconstruction quality. Additionally, both our method and TetriRF offer rendering times that are at least twice as fast as the ReRF and JointRF.
Tab. \ref{time} presents a detailed analysis of the computational complexity of our VRVVC model on the ReRF dataset, showing that VRVVC offers excellent computational efficiency and serves as an effective solution for volumetric video compression.

\begin{table}[h]
\setlength{\tabcolsep}{9pt}
\renewcommand{\arraystretch}{1.1}
\centering
\scalebox{1}{
\begin{tabular}{cccc}
\hline
Train(min) & Render(s) & Encode(s) & Decode(s) \\ \hline
2.6       & 0.13     & 1.23      & 0.88      \\ \hline
\end{tabular}
}
\vspace{-2mm}
\caption{Complexity analysis results of VRVVC.}
\label{time}
\vspace{-2mm}
\end{table}

\begin{table}[t]
\renewcommand\arraystretch{0.94}
\begin{tabular}{c|c|c|c}
\hline
\multirow{2}{*}{Dataset}       & \multirow{2}{*}{Method} & \multicolumn{1}{c|}{Training View}                                                                            & \multicolumn{1}{c}{Testing View}                                                                             \\ \cline{3-4} 
                               &                          & BDBR(\%) $\downarrow$  & BDBR(\%) $\downarrow$\\ \hline
\multirow{3}{*}{ReRF}          & ReRF                                                                   & 424.97                                                                                            & 346.58                                              \\
& JointRF                                                                   & 127.70                                                                                      & 90.33                                              \\
                               & Ours                                                              & \textbf{-46.25}                                                                                           & \textbf{-48.27}                                               \\ \hline
\multirow{3}{*}{\makecell{DNA-\\Rendering}} &  ReRF                                                                & 177.71                                                                                        & 103.56                                              \\
& JointRF                                                                   & 2.99                                                                                           & 0.32                                              \\
                               & Ours                                                                & \textbf{-81.86}                                                                                          & \textbf{-83.51}                                               \\ \hline
\end{tabular}
\vspace{-2mm}
\caption{The BDBR results of our VRVVC, ReRF and JointRF when compared with TeTriRF on different datasets.}
\label{table:BDBR}
\vspace{-1em}
\end{table}

\begin{figure}[t]
  \centering
  \begin{subfigure}[b]{0.49\linewidth}
    \includegraphics[width=\linewidth]{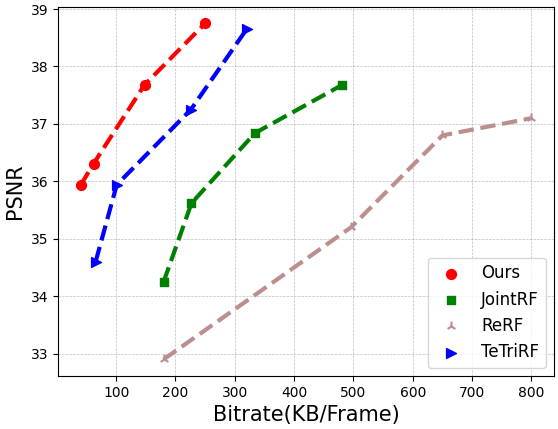}
    \caption{ReRF (Train) }
    \label{fig:sub1}
  \end{subfigure}
  \hfill
  \begin{subfigure}[b]{0.49\linewidth}
    \includegraphics[width=\linewidth]{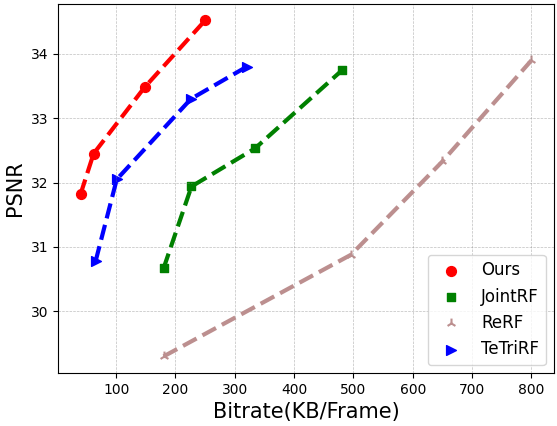}
    \caption{ReRF (Test)}
    \label{fig:sub2}
  \end{subfigure}
  
  \begin{subfigure}[b]{0.49\linewidth}
    \includegraphics[width=\linewidth]{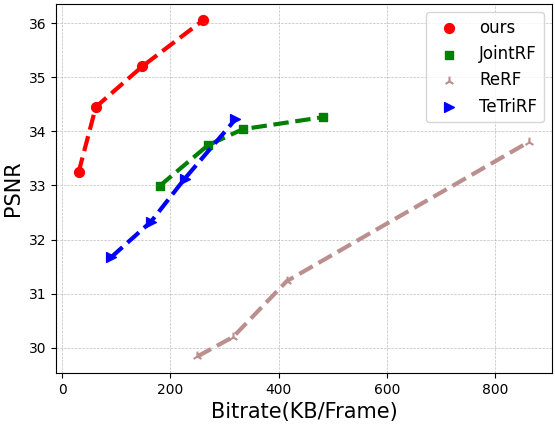}
    \caption{DNA-Rendering (Train)}
    \label{fig:sub3}
  \end{subfigure}
  \hfill
  \begin{subfigure}[b]{0.49\linewidth}
    \includegraphics[width=\linewidth]{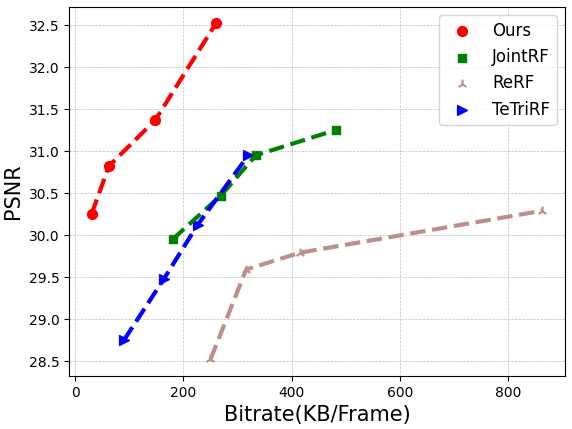}
    \caption{DNA-Rendering (Test)}
    \label{fig:sub4}
  \end{subfigure}
  \vspace{-2mm}
  \caption{ The RD performance comparison results on the ReRF and DNA-Rendering datasets. }
  \label{pic:rd}
  \vspace{-2mm}
\end{figure}

\begin{figure}[t]
  \centering
  \begin{subfigure}[b]{0.49\linewidth}
    \includegraphics[width=\linewidth]{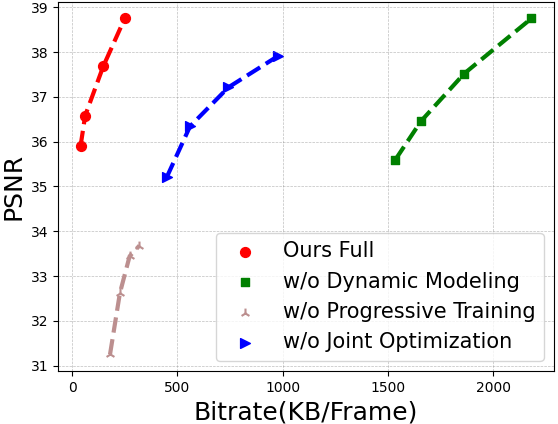}
    \caption{Training View}
    \label{fig:sub5}
  \end{subfigure}
  \hfill
  \begin{subfigure}[b]{0.49\linewidth}
    \includegraphics[width=\linewidth]{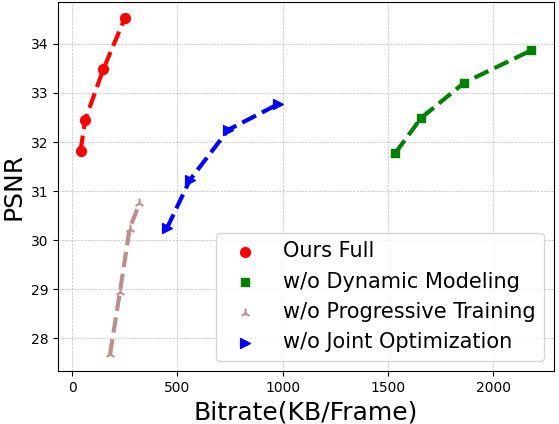}
    \caption{Testing View}
    \label{fig:sub6}
  \end{subfigure}
  \vspace{-2mm}
  \caption{RD curves. This figure illustrates the efficiency of various components within our method.}
  \label{pic:rd3}
  \vspace{-1em}
\end{figure}

The RD performance of our VRVVC compared with ReRF, TeTriRF, and JointRF is presented in Tab. \ref{table:BDBR}. Notably, our VRVVC consistently outperforms these methods in terms of RD performance. For instance, compared to TeTriRF, our method achieves average BDBR reductions of \textbf{-46.25\%} for training views and \textbf{-48.27\%} for testing views on the ReRF dataset. Similarly, on the DNA-Rendering dataset, we observe average BDBR savings of \textbf{-81.86\%} for training views and \textbf{-83.51\%} for testing views. The RD curves, shown in Fig. \ref{pic:rd}, further illustrate that our VRVVC achieves superior RD performance across a wide range of bitrates. It is worth noting that while JointRF requires training multiple fixed-bitrate models to achieve different rate-distortion trade-offs, our method provides a broader range of RD performance with just a single model, offering greater flexibility and efficiency.

\begin{figure}[ht]
    \centering
    \includegraphics[width=\linewidth]{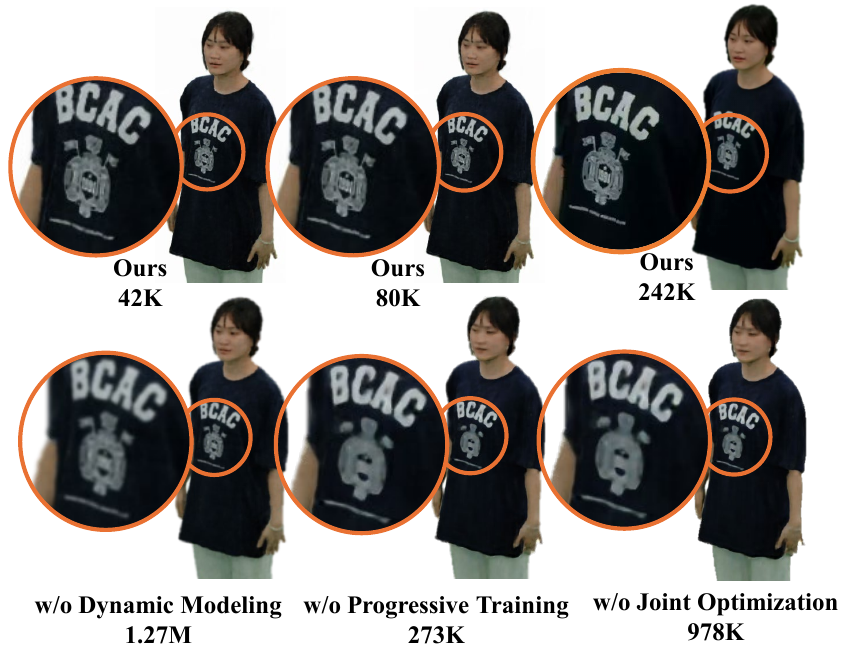}
    \vspace{-6mm}
    \caption{Qualitative results of complete VRVVC and its variants. Excluding any module results in lower reconstruction quality and an increase in bitrate.}
    \label{fig:ablation}
\end{figure}
\vspace{-2mm}
\subsection{Ablation Studies}
We perform three ablation studies to evaluate the effectiveness of residual dynamic modeling, progressive training, and joint optimization by disabling each component individually during training. In the first study, we model volumetric video frame by frame without applying residual dynamic modeling. In the second, we skip the initial stage and train the entire framework directly. In the final study, we train the residual representation and entropy model separately instead of optimizing them jointly.

The results of the ablation studies can be seen in Fig. \ref{pic:rd3}. It shows that disabling either residual dynamic modeling or progressive training leads to an increase in bitrate, underscoring the effectiveness of these modules. Additionally, joint optimization produces temporally consistent and low-entropy 4D sequential representations, which are more efficiently compressed, thereby significantly enhancing RD performance. Fig. \ref{fig:ablation} presents a qualitative comparison of the complete VRVVC at different bitrates against its variants. These findings highlight the advantages of our residual dynamic modeling, progressive training, and joint optimization strategy in volumetric video compression.

\vspace{-4mm}
\section{Conclusion}

In this paper, we present a novel variable-rate compression framework tailored for NeRF-based volumetric video. Our tri-plane residual representation in VRVVC is compact and compression-friendly, effectively reducing spatio-temporal redundancy between frames in a sequential manner. Our residual representation compression scheme employs an implicit entropy model coupled with RD tradeoff parameters to enable variable bitrates. Our end-to-end training strategy jointly optimizes both representation and compression, significantly improving compression performance. Experimental results demonstrate that VRVVC not only achieves a wide range of variable bitrates within a single model but also surpasses state-of-the-art fixed-rate methods, greatly advancing the transmission capabilities of volumetric video.
\vspace{-6mm}
\section*{Acknowledgements}
This work was supported by National Natural Science Foundation of China (62271308), STCSM (24ZR1432000,24511106902, 22511105700, 22DZ2229005), 111 plan
(BP0719010), Open Project of National Key Laboratory of China (23Z670104657) and State Key Laboratory of UHD Video and Audio Production and Presentation.
\clearpage
\bibliography{main.bbl}
\end{document}